\documentclass[conference]{IEEEtran}
\IEEEoverridecommandlockouts
\usepackage{cite}
\usepackage{amsmath,amsfonts}
\usepackage{algorithmic}
\usepackage{algorithm}
\usepackage[dvipsnames]{xcolor}
\usepackage{array}
\usepackage{cite}
\usepackage[caption=false,font=normalsize,labelfont=sf,textfont=sf]{subfig}
\usepackage{textcomp}
\usepackage{stfloats}
\usepackage{url}
\usepackage{verbatim}
\usepackage{graphicx}
\usepackage{xcolor}
\usepackage{cite}
\usepackage{algorithmic}
\usepackage{algorithm}
\usepackage{multirow}
\usepackage{enumitem}
\usepackage{silence}
\usepackage{flushend}
\usepackage{amsmath,amssymb,amsfonts}
\usepackage{algorithmic}
\usepackage{graphicx}
\usepackage{textcomp}
\usepackage{xcolor}
\DeclareMathOperator{\ASR}{ASR}
\DeclareMathOperator{\CAD}{CAD}
\DeclareMathOperator{\acc}{acc}

\def\BibTeX{{\rm B\kern-.05em{\sc i\kern-.025em b}\kern-.08em
    T\kern-.1667em\lower.7ex\hbox{E}\kern-.125emX}}
\begin{document}

\title{Genetic Algorithm-Based Dynamic Backdoor Attack on Federated Learning-Based Network Traffic Classification
}

\author{\IEEEauthorblockN{
Mahmoud Nazzal\IEEEauthorrefmark{1},
Nura Aljaafari\IEEEauthorrefmark{4},
Ahmed Sawalmeh\IEEEauthorrefmark{6}, 
Abdallah Khreishah\IEEEauthorrefmark{1},\\
Muhammad Anan\IEEEauthorrefmark{6},
Abdulelah Algosaibi\IEEEauthorrefmark{4},
Mohammed Alnaeem\IEEEauthorrefmark{4},
Adel Aldalbahi\IEEEauthorrefmark{4},
Abdulaziz Alhumam\IEEEauthorrefmark{4},\\
Conrado P. Vizcarra\IEEEauthorrefmark{4}, and
Shadan Alhamed\IEEEauthorrefmark{4}
}
\IEEEauthorblockA{\IEEEauthorrefmark{1} New Jersey Institute of Technology, Newark, NJ 07102, USA}
\IEEEauthorblockA{\IEEEauthorrefmark{4} 
King Faisal University, Al Hofuf 31982, KSA
}
\IEEEauthorblockA{\IEEEauthorrefmark{6} 
Alfaisal University, Riyadh 11533, KSA}
\\
E-mails: 
mahmoud.nazzal@ieee.org,
naaljaafari@kfu.edu.sa, 
asawalmeh@alfaisal.edu,
abdallah@njit.edu\\
manan@alfaisal.edu,
\{aaalgosaibi,
naeem,
aaldalbahi,
aahumam,
cvizcarra,
ssalhamed\}@kfu.edu.sa
}

\maketitle

\begin{abstract}
Federated learning enables multiple clients to collaboratively contribute to the learning of a global model orchestrated by a central server. This learning scheme promotes clients' data privacy and requires reduced communication overheads. In an application like network traffic classification, this helps hide the network vulnerabilities and weakness points. However, federated learning is susceptible to backdoor attacks, in which adversaries inject manipulated model updates into the global model. These updates inject a salient functionality in the global model that can be launched with specific input patterns. Nonetheless, the vulnerability of network traffic classification models based on federated learning to these attacks remains unexplored. In this paper, we propose GABAttack, a novel \underline{g}enetic \underline{a}lgorithm-based \underline{b}ackdoor \underline{attack} against federated learning for network traffic classification. GABAttack utilizes a genetic algorithm to optimize the values and locations of backdoor trigger patterns, ensuring a better fit with the input and the model. This input-tailored dynamic attack is promising for improved attack evasiveness while being effective. Extensive experiments conducted over real-world network datasets validate the success of the proposed GABAttack in various situations while maintaining almost invisible activity. This research serves as an alarming call for network security experts and practitioners to develop robust defense measures against such attacks.
\end{abstract}


\section{Introduction}
\par Network traffic classification (NTC) categorizes network traffic observations based on their features and characteristics to infer certain properties. NTC is an integral part of network management commonly employed by network administrators and service providers. A direct outcome of NTC is giving an insight into the types of activities, applications, and protocols used network-wide. This information is essential for optimizing the network resources and identifying any potential threats or malicious actions. Classical NTC approaches include port-based classification, deep packet inspection \cite{song2020software}, and statistical classification \cite{su2005extended}. Similar to the case with many other application areas, machine learning (ML) models have been successfully utilized in NTC due to their outstanding performances as data-driven approaches \cite{alshammari2015identification}. 

\par Conventional ML models used in NTC share a common limitation; depending on manually crafted features that typically require experts' knowledge, practice, and time. As an example, \cite{zhongsheng2020retracted} utilizes support vector machines (SVM) operating on 250 network flow features proposed in \cite{moore2013discriminators}. A common drawback of these approaches is also requiring a reasonable size of training data to train the model.

\par To resolve the limitations of standard ML models for NTC, recent research considers a growing interest in federated learning (FL) \cite{mothukuri2021survey,banabilah2022federated} as a framework for model training. FL aggregates user-end ML model contributions to obtain a global model. Since FL allows local training on the client side, it achieves two main advantages; promoting the client's privacy, and saving the network bandwidth as only model coefficients are communicated \cite{kairouz2019advances}. These are attractive features for an application like NTC since NTC data can easily reveal network vulnerabilities and weaknesses. Accordingly, several works have recently considered FL for NTC
\cite{mun2020internet,he2021edge,peng2021federated,guo2022feat,sun2022traffic}. 

\par Despite the attractive advantages of FL, it is widely believed to be inherently vulnerable to adversarial and poisoning attacks \cite{kairouz2019advances} similar to the case of virtually all ML settings due to their data dependency. In a security-critical application like NTC, malicious actors have strong incentives to attack NTC models to tweak their cyber attacks \cite{ahmed2017poster}. A key incentive is to limit their cyber attacks to be within legitimate network traffic \cite{usama2019adversarial} thereby bypassing NTC-based network intrusion detection and firewalls. This highlights the importance of understanding what creates this vulnerability and developing efficient countermeasures accordingly. 

\par Even though FL keeps data at clients, FL NTC models can enable multiple clients to access model parameters and thus empower malicious intervention. While recent literature has a few works on establishing the vulnerability of NTC models to adversarial attacks, their susceptibility to backdoor attacks is yet studied. Particularly, the vulnerability of the recent FL-based NTC approach is not addressed. A backdoor injects a salient functionality into a target model. This functionality is only activated if a certain trigger pattern is available in a test input. To keep the attack evasive, the target model should behave normally with benign inputs \cite{pang2020tale}. While various existing backdoor attacks perform relatively well on effectiveness, evasiveness is a more challenging goal \cite{doan2020februus,liu2019abs}.
 
\textbf{Contributions} Motivated by the above discussion; we present the following contributions.
\begin{itemize}[leftmargin=*]
\item Establishing the vulnerability of FL-based NTC models to backdoor attacks characterized by specially crafted triggers injected in training and activated in inference.
\item GABAttack: a new algorithm for backdoor attack against FL optimizing the locations and values of trigger patterns in network traffic data based on genetic algorithm optimization. GABAttack offers attack transferability and produces input-tailored dynamic triggers enhancing attack evasiveness.
\item A comprehensive set of experiments to investigate the performance of GABAttack in real-world NTC attack in terms of effectiveness evasiveness.
\end{itemize}

\begin{figure}[thb]
\centering
\resizebox{0.99\columnwidth}{!}{
\begin{tabular}{cc}
\includegraphics[scale=0.34]{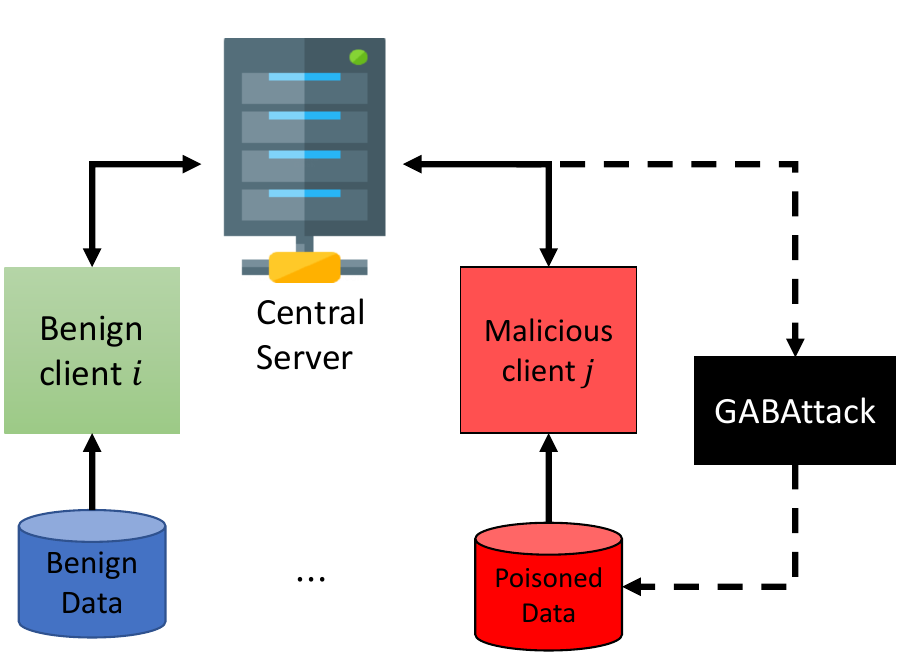}
&
\includegraphics[scale=0.44]{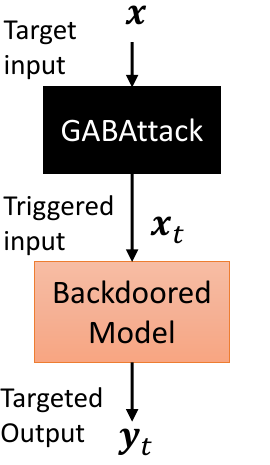}
\\
(a)
&
(b)
\end{tabular}}
\caption{GABAttack: BD pattern injection in (a) where the adversary modifies the training data of compromised clients to inject a backdoor functionality in the model updates submitted to the server. Attack launching in (b).}
\label{system_overview}
\end{figure}

\section{Background} 
\label{Section2}

\par \textbf{FL-based NTC} FL \cite{mcmahan2017communication} is a setting for distributed ML where $n$ clients collaboratively learn a global model $\phi$. In a training round $t \in\{1, \ldots, T\}$, each client $i \in\{1, \ldots, k\}$ trains a model $\phi_i$ on its local data $D_i$ with based on the previous global model $\phi_{t-1}$. The client model (or its update over the previous global model) is then communicated to the central server. This server aggregates all client model updates to obtain an updated global model, and the process is repeated in the next FL round. Since FL is based on communicating model coefficient updates rather than data points, it naturally promotes data privacy while substantially reducing the communication overhead compared to standard distributed learning. Thus, FL has gained popularity in a wide spectrum of applications ranging from healthcare \cite{chen2020fedhealth} to autonomous driving \cite{liang2019federated} and, more recently, NTC \cite{mun2020internet,he2021edge,peng2021federated,guo2022feat,sun2022traffic}. In an NTC context, FL is used for data privacy reasons \cite{mun2020internet}. FL is enhanced with an attention mechanism for better client model aggregation in \cite{zhu2022attention}. 
Other works \cite{he2021edge,sanchez2023robust} use FL-based NTC for device identification. These works assume honest clients and overlook the possibility of having some clients potentially compromised by malicious actors as depicted in Fig. \ref{system_overview}.

\textbf{Genetic algorithm} (GA) \cite{banzhaf1998genetic} is a meta-heuristic search algorithm that has been widely used in many application areas. The idea behind GA is based on alternating between the generation of new candidate solutions and selecting the best among them. These operations are inspired by the processes of evolution and natural selection in evolutionary theory. The deployment of GA requires first encoding data (either inputs or solutions) into \textit{chromosomes}. A chromosome represents the information of a given solution. GA then starts from an arbitrary set of initial candidate solutions referred to as the parents. Then, the parent set is iteratively refined by yielding new solutions referred to as the offspring. Fundamentally, GA uses the processes of mutation and crossover to obtain offspring from parents. Next, in each iteration, the set of parents is updated by the inclusion of specific members of the offspring, replacing specific elements in the parent set. This is done based on a certain fitness function. For the sake of diversity and exploration, GA balances between greedily selecting the best candidates and keeping a few ``bad'' ones.

\begin{figure}[htb]
\centering
\resizebox{0.99\columnwidth}{!}{
\includegraphics{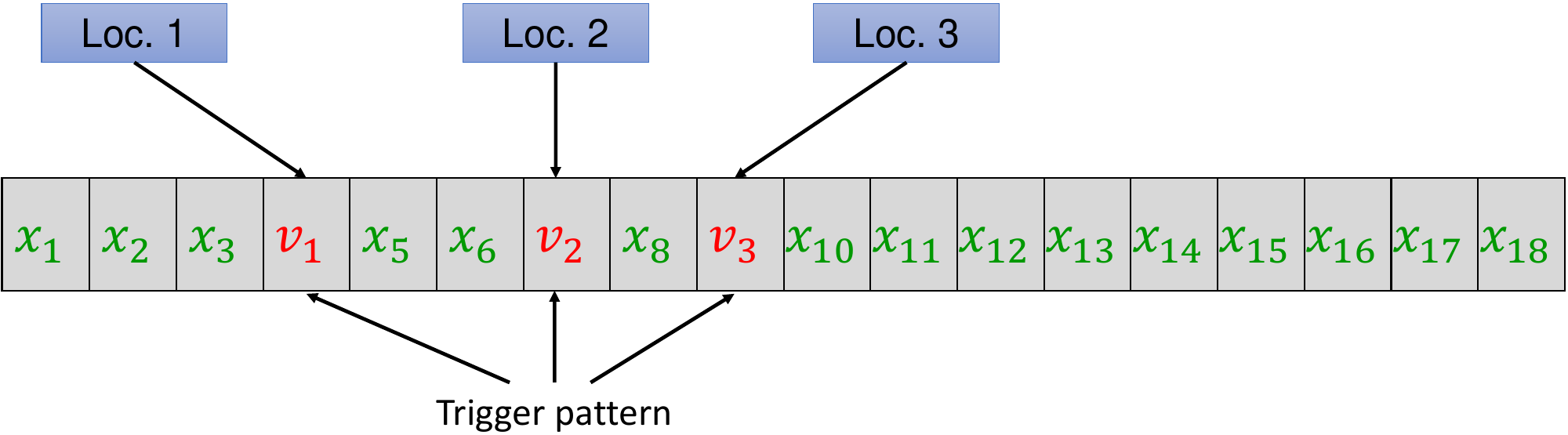}}\caption{Chromosome construction and encoding, where $x_1$ though $x_{18}$ are the elements of the target feature input and $v_1$ though $v_3$ are trigger values at locations $Loc 1$ through $Loc 3$.}
\label{chrom_encoding}
\end{figure}

\begin{figure*}[htb]
\centering
\resizebox{0.99\textwidth}{!}{
\includegraphics{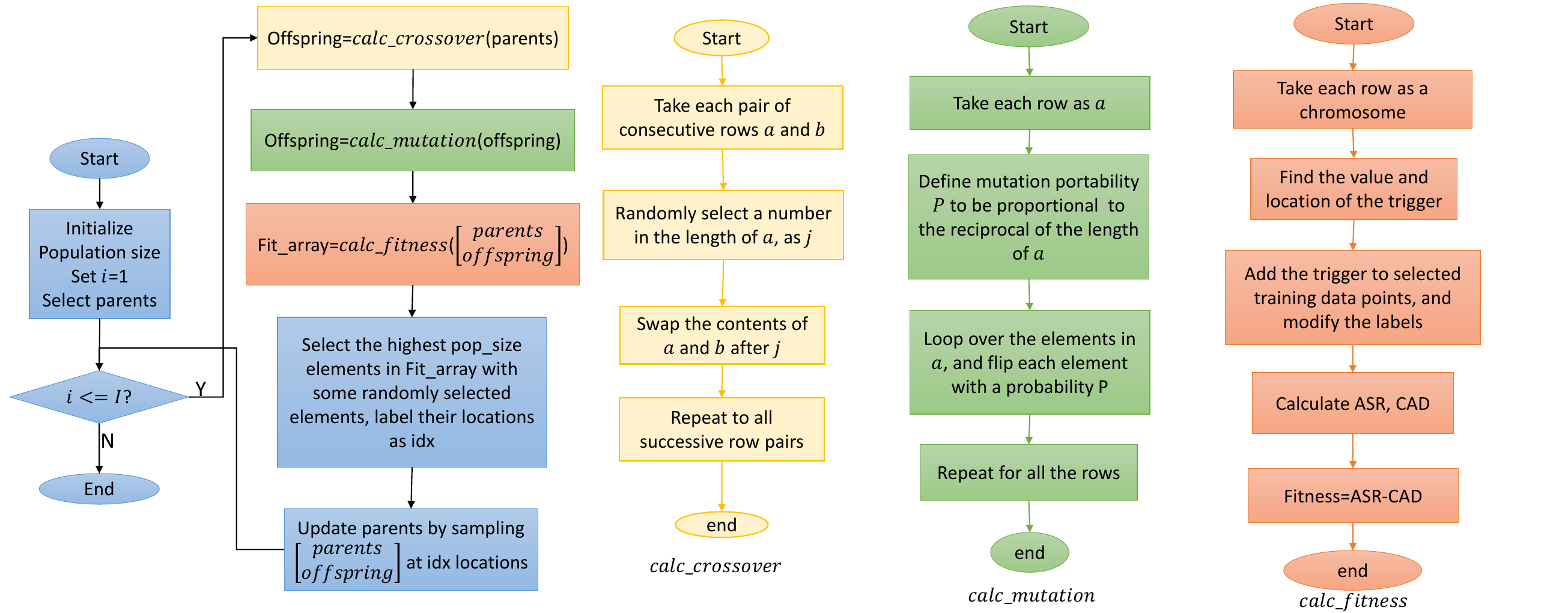}
}
\caption{GABAttack in flowchart description.}
\label{flowchart_proposed}
\end{figure*}

\section{Threat Model}
\par We consider an FL-based NTC system \cite{mun2020internet,he2021edge,peng2021federated,guo2022feat,sun2022traffic} composed of $N$ clients orchestrated by a server as shown in  Fig. \ref{system_overview}(a). FL is operated in rounds where in each round $t$, the server randomly selects $k=C.N, C \leq 1$ clients to participate in the training process. We characterize the threat model in terms of the adversary's objectives, knowledge, and capabilities. The objective of the adversary is to craft \textit{effective} and \textit{evasive} (salient) backdoor attacks on the target model. The adversary is assumed to possess a few of the FL clients and to use them to inject the backdoor during training and launch it during testing times. So, the adversary controls the training data and training operations for its clients only. Specifically, the adversary can manipulate the training data points and their labels at these clients as shown in  Fig. \ref{system_overview}(a). Besides, it knows the initial model broadcast to them by the server. For launching the attack as shown in  Fig. \ref{system_overview}(b), the adversary needs to know what network data to target. We assume that the adversary can observe legitimate packets by ``sniffing the network'' and therefore record the feature values of legitimate traffic as commonly assumed in adversarial attack works against NTC models \cite{ahmed2017poster,wang2014man}.

\section{The proposed GABAttack}
\label{Section3}
\par Fig. \ref{system_overview} illustrates the workflow of the GABAttack. First, GABAttack is used to inject a backdoor functionality in the target model during training (part a). Then, it is used to launch an attack during the run time of the model (part b). The goals GABAttack aims at when generating backdoor patterns are effectiveness and evasiveness, as represented below. 
\[
\phi(x)=\left\{
\begin{array}{rl}
\phi(x_t)=y_t& effectiveness\\
\phi(x)=\phi_0(x)& evasiveness
\end{array}
\right.
\]
\noindent where $\phi$ is a backdoored model corresponding to a being model $\phi_0$, $x$ is a benign input, $x_t$ is a triggered input, and $y_t$ is the targeted label of $x_t$. According, the backdoor attack problem can be formulated as follows.
\begin{equation}
    \arg \min _{\boldsymbol{\phi}^{\star}, D_m} \mathcal{L}_M\left(D_m ; \boldsymbol{\phi}^{\star}\right)+\rho\left\|\Delta_m^{t+1}-\bar{\Delta}_{\text {ben }}^t\right\|_2
\end{equation}
\noindent where $D_m$ denotes the data of a malicious client, $\rho$ is a hyperparameter, $\mathcal{L}_M$ is Cross-entropy loss for main task, and $\bar{\Delta}_{\text{ben }}^t$ denotes benign clients' averaged model updates. In this paper, $\bar{\Delta}_{\text {ben }}^t$ is estimated using a shared global model. We assume that the aggregated global model is similar to the benign client's local model as the shared global model converges to a point with a high testing accuracy. It should be noted that the addition of a regularization term is not sufficient to ensure that the malicious weight update is close to that of the benign agents since there could be multiple local minima with similar loss values.

\par GABAttack employs a GA-based approach for optimizing both the values of the added trigger elements and their placement in the data. This requires first developing a chromosome encoding of the GA. Fig. \ref{chrom_encoding} shows the proposed chromosome encoding process along with the backdoor pattern placement. A direct measure of an attack's effectiveness is the attack success rate (ASR) defines as follows.
\begin{equation}
\ASR=\frac{I(\phi({x_t})=y_t)}{num}
\end{equation}
\noindent where $x_t$ is a triggered input, $y_t$ is the targeted model outcome, $num$ is the total number of attacked inputs, and $I$ is an indicator function. 
\par As for evasiveness, one can quantify it as the drop in model accuracy working on benign data after having a trigger functionality. Thus, it can be calculated as 
\begin{equation}
\CAD=\acc(\phi_0(X),Y)-\acc(\phi(X),Y)
\end{equation}
\noindent where $\acc$ is an accuracy function, $X$ is a set of benign training data with $Y$ true labels, $\phi_0$ is a benign trained model, and $\phi$ is its backdoored version. Accordingly, we define the following GA fitness function to incorporate ASR and CAD.
\begin{equation}
    f=\ASR-\gamma \CAD
\end{equation}
\noindent where $\gamma$ balances the trade-off between ASR and CAD.

\par The main steps of GABAttack are outlined in the flowchart of Fig. \ref{flowchart_proposed}. Along with maintaining attack effectiveness and evasiveness, there are several advantages of GA in optimizing the triggers. First, GA is naturally a global optimization algorithm that is likely to generate globally optimized outcomes. Second, it allows searching over the set of possible candidates without the need for establishing this set ahead of time.  

\section{Experiments}
\label{Section4}
\par We examine the performance of GABAttack first in a centralized setting assuming white-box model access and then in the intended usage in an FL setting with the same datasets and models.

\subsection{Experimental setup}
\par We consider the widely used Moore \cite{moore2013discriminators} dataset in all the experiments. This dataset uses 216 features used in training and inference. As for the classes, the top six classes are assumed; \textit{WWW, MAIL, FTP-DATA, FTP-CONTROL, DATABASE, SERVICES,} and \textit{ATTACK}. As for the simulation platform, experiments are run over Google's Colab with Keras 2.8.0 with Tensorflow Backend, and Python 3.7.13. 

\par For the centralized setting, we assume the poisoning occurs on the data the model directly uses for training, where it is combined with clean data with the correct label. The model is trained for 3 epochs, with a batch size of 100 and a learning rate of 0.001. For the FL experiments, the data is distributed across the clients following an IID nature. In each FL round, 10 clients are assumed to be participating in the model training. The benign clients are assumed to follow the traditional procedure of training the model and that they are training the model over 3 epochs with a batch size of 100 data points. Malicious clients perform two operations; first, they run GABAttack on their whole training data to get the best trigger values and locations. Then only 50\% at maximum is cut from their training data and poisoned with the trigger. The second operation is model training, in which the malicious clients combine the clean data and the poisoned data and then use the combined data to train their models. Following similar training parameters to the benign clients, it is assumed that each malicious client trains its model for 3 epochs, with a batch size of 100, and a learning rate of 0.001. 

\par The model accuracy and ASR are recorded to measure the attack's success. In the FL setting, it is reported for malicious clients, and the same metrics are reported for the global model. Furthermore, the global model accuracy on the benign data is recorded at the end of each FL round. However, since this is a dynamic attack, where each malicious client has its trigger and poisons the data in a different location, this presents a challenge of how to measure the attack success on the global model. Thus, the attack success in the FL is measured in the global model for each trigger in an attack round. 

\begin{figure}[b]
\centering
\resizebox{0.95\columnwidth}{!}{
\begin{tabular}{ccc}
\includegraphics[width=10cm,height=10cm]{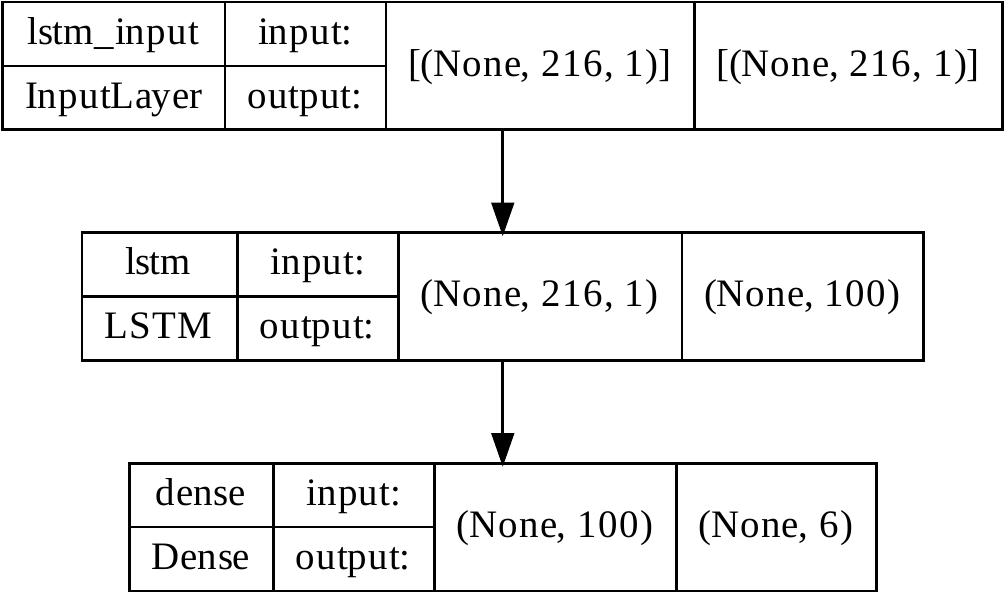}
&
\includegraphics[width=10cm,height=10cm]{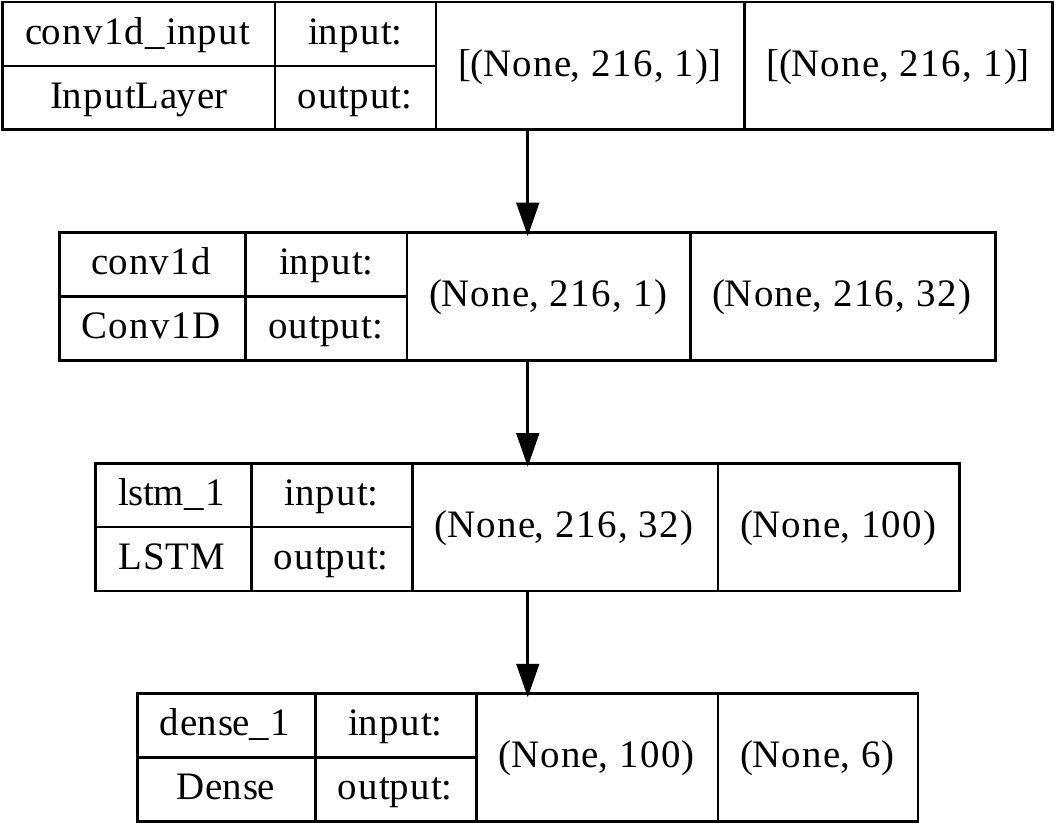}
&
\includegraphics[width=10cm,height=10cm]{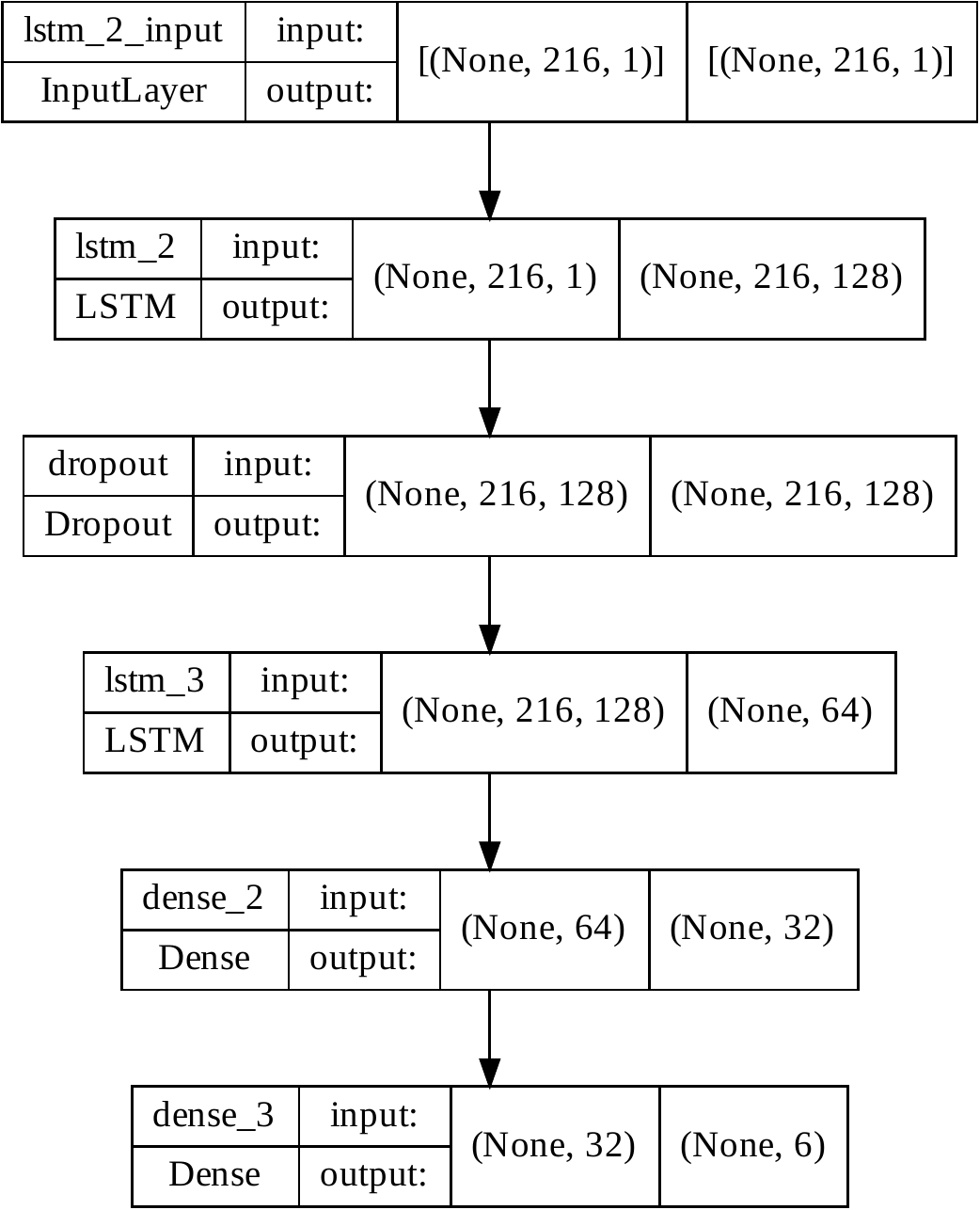}
\\
\Huge{(a)}
&
\Huge{(b)}
&
\Huge{(C)}
\end{tabular}}
\caption{The architectures of the $Simple_{LSTM}$, $Convolutional_{LSTM}$, and $Complex_{LSTM}$ models in (a), (b), and (c), respectively.}
\label{models_considered} 
\end{figure}

\par \textbf{Classification models} We use the following three models. The first model, shown in Fig. \ref{models_considered}(a) is a neural network with one long short-term memory (LSTM) layer of 100 hidden units, following a fully connected layer. The second model is a neural network that contains a layer of 1D convolution layer, then an LSTM layer with 100 units, followed by a fully connected layer. The structure of $Convolutional_{LSTM}$ is shown in Fig. \ref{models_considered}(b). The third model as shown in Fig. \ref{models_considered}(c), is a neural network that has an LSTM layer with 100 units, then another LSTM layer with 64 units. The model includes 2 fully-connected layers. The three models are created incrementally in terms of the depth and the number of parameters in each model, i.e., their complexity.
\subsection{Experiments}

\begin{figure}[htb]
\centering
\resizebox{0.95\columnwidth}{!}{
\includegraphics[scale=0.44]{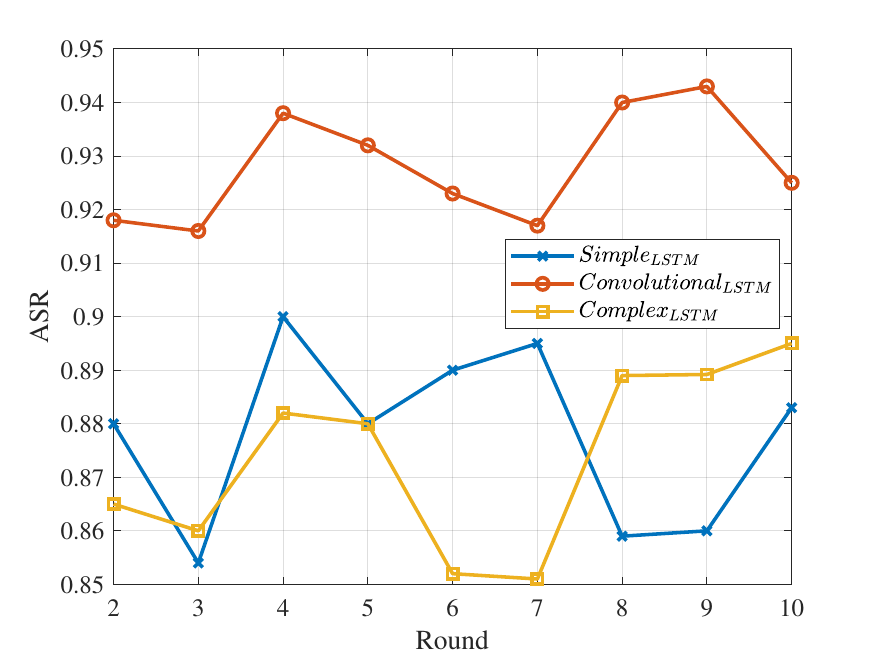}}\caption{Attack performance when the percentage of adversary clients is 20\%.}
\label{Fig_5_new}
\end{figure}

\par The first point that is investigated is the attack success over different models, that is, testing if the attack can succeed in deep and complex models and the attack transferability to complex models. It is worth noting that several experiments are conducted using traditional ML, such as SVM, and decision trees, but we do not present their results as they perform poorly on the main classification task, and are thus of low practical interest. For this experiment, in the FL setting, it is assumed that 10\% of the clients participating in the pool are malicious and the attack is initiated on the fourth round. It is also assumed that adversary clients run independently. 

\par Table. \ref{Centralized_model_results} demonstrates the centralized model accuracy on the aforementioned models as well as the ASR. For the training setting here, it is assumed that the model is trained from scratch. It is seen that the attack succeeds in the cases without compromising the model accuracy on the benign data. It is noted that the models considered show better results on the data than most models presented in the literature. 

\begin{table}[ht]
\centering
\caption{Centralized model results.}\label{Centralized_model_results}
\resizebox{0.9\columnwidth}{!}{
\begin{tabular}{|c|c|c|c|}
\hline
 & $Simple_{LSTM}$ & $Convolutional_{LSTM}$ & $Complex_{LSTM}$ \\ \hline
Accuracy & 93.16\%  & 95.75\%  & 95.59\%  \\ \hline
ASR      & 99.\%  & 100.0\%  & 100.0\%\\ \hline
\end{tabular}}
\end{table}

\begin{figure}[htb]
\centering
\resizebox{0.95\columnwidth}{!}{
\includegraphics[scale=0.44]{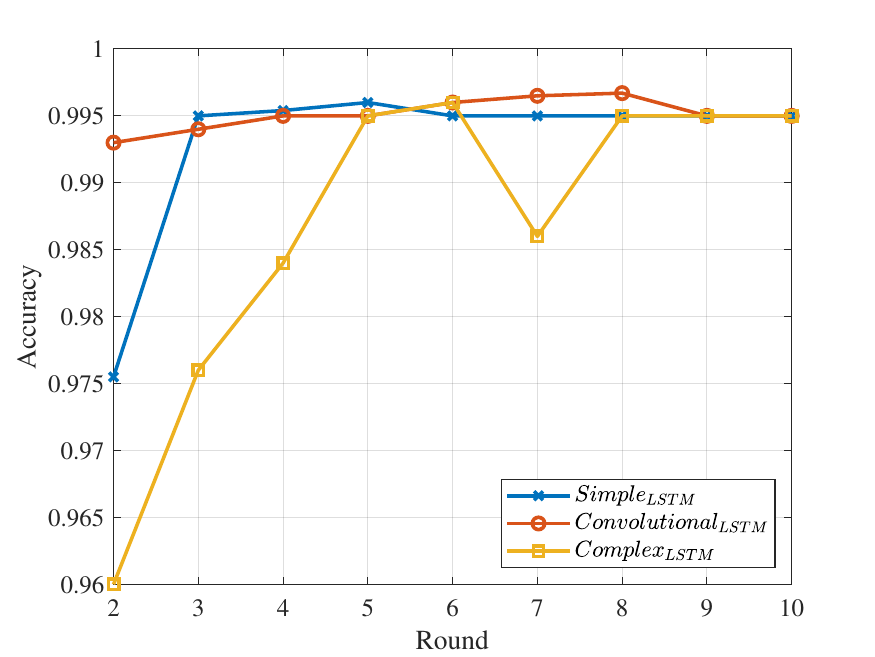}}\caption{Attack performance when the percentage of adversary clients is 50\%.}
\label{Fig_6_new}
\end{figure}

\par Next, we examine the attack performance metrics in the FL setting. Fig \ref{Fig_5_new} shows the ASR versus FL round with the three models considered. It can be seen that the ASR succes is cummunceurate with model's compeleixty. This can be interprested in view of the fact that model complexity opromises for accomodating for both the main task and the backdoor functionality. Moreover, the attack is geneally highlyasuccessful for the three models cosnidered. As for attack evasiveness, Fig. \ref{Fig_6_new} shows the glodbal model accuracy with the three models. It can be seen that the most complex model  exhibits less model acucracy in the first inital rounds. However, the model accuracy is maiantanied for the three models. This result assures the evasiveness of the attack.

\par As for the centralized setting, the ASR is at its highest on the attack round, which is around 99.98\% on average, but it slightly decreases on the following rounds, reporting 98.73\% and 98.20\%, respectively. The results presented in this section show that the GABAttack's attack is successful in terms of both attack's effectiveness and evasiveness. 

\section{Related Work}

\par \textbf{Backdoor Attack on FL Models} In backdoor attacks, the adversary manipulates the local models of compromised clients to obtain poisoned models to be then aggregated into the global model. There are many works of backdoor attacks on FL. Several recent works demonstrate the susceptibility of the FL model to backdoor attacks. These works assume that some FL clients are compromised by the adversary and under its control \cite{shen2016auror, xie2020dba}. Some works concentrate on scaling the impact of malicious clients’ contributions to the global model in what is known as model replacement attacks \cite{ bagdasaryan2020backdoor }. Other works focus on keeping the attack as stealthy as possible \cite{wang2020attack}.  

\par \textbf{Adversarial attack on NTC models} is still in its early stages. The current research body in this area focuses mainly on developing adversarial examples to manipulate NTC outcomes. Along this line, \cite{verma2018network} employs the well-known Carlini and Wanger method \cite{carlini2017towards} to generate such adversarial examples. Next, \cite{usama2018adversarial} investigates a range of established adversarial attacks in targeting NTC models thus demonstrating their vulnerability. Subsequently, \cite{usama2019black} leverages mutual information to identify the optimal features to perturb to manipulate NTC classification. It can be seen that this research body focuses on interference-time evasive attacks and overlooks backdoor attacks. Besides, to the best of our knowledge, there are no attacks on FL-based NTC.

\section{Conclusion}
\label{Section5}
In this paper, we propose GABAttack, a new backdoor attack against FL-based NTC models. The proposed GABAttack uses a genetic algorithm as a means for tuning the values and locations of backdoor trigger patterns to best fit the input and the model if known. Thus, this is an input-tailored dynamic attack promising improved attack evasiveness. Extensive experiments conducted over real-world NTC data with varying model complexities validate the success of the proposed GABAttack in terms of attack effectiveness and evasiveness. This work establishes the vulnerability of FL-based NTC models to backdoor attacks and calls for devising viable defense measures against such attacks. Future work will include the design of a coordinated attack across adversary-compromised clients and better ways of balancing the effectiveness-evasiveness trade-offs in the attack. 

\section*{Acknowledgement}
\par The authors extend their appreciation to the Deputyship
for Research \& Innovation, Ministry of Education in Saudi
Arabia for funding this research work through the project
number 1120. 

\bibliography{refs.bib}
\bibliographystyle{IEEEtran}
\end{document}